\documentclass[iop,apjl]{emulateapj}
\usepackage{amssymb}
\usepackage{amsmath}
\usepackage{graphicx}

\begin{document}
\journalinfo{ApJ Letters, in press}

\shortauthors{Cyganowski et al.}

\title{The Protocluster G18.67+0.03: A Test Case for Class I CH$_{3}$OH Masers as Evolutionary Indicators for Massive Star Formation}
\author{C.J. Cyganowski\altaffilmark{1,6}, C.L. Brogan\altaffilmark{2},
  T.R. Hunter\altaffilmark{2}, Q. Zhang\altaffilmark{1},
  R.K. Friesen\altaffilmark{2,3}, R. Indebetouw\altaffilmark{2,4}, C.J. Chandler\altaffilmark{5}}

\email{ccyganowski@cfa.harvard.edu}

\altaffiltext{1}{Harvard-Smithsonian Center for Astrophysics, Cambridge, MA 02138}
\altaffiltext{2}{NRAO, 520 Edgemont Rd, Charlottesville, VA 22903}
\altaffiltext{3}{Dunlap Institute for Astronomy and Astrophysics, University of Toronto, Toronto ON Canada M5S 3H4}
\altaffiltext{4}{Department of Astronomy, University of Virginia, Charlottesville, VA 22903}
\altaffiltext{5}{NRAO, 1003 Lopezville Road, Socorro, NM 87801}
\altaffiltext{6}{NSF Astronomy and Astrophysics Postdoctoral Fellow}

\begin{abstract}

We present high angular resolution Submillimeter Array (SMA) and
Karl G. Jansky Very Large Array (VLA) observations of the massive protocluster
G18.67+0.03.  Previously targeted in maser surveys of GLIMPSE Extended
Green Objects (EGOs), this cluster contains three Class I CH$_{3}$OH
maser sources, providing a unique opportunity to test the proposed
role of Class I masers as evolutionary indicators for massive star
formation.  The millimeter observations reveal bipolar molecular
outflows, traced by $^{13}$CO(2-1) emission, associated with all three
Class I maser sources.  Two of these sources (including the EGO) are
also associated with 6.7 GHz Class II CH$_{3}$OH masers; the Class II
masers are coincident with millimeter continuum cores that exhibit hot
core line emission and drive active outflows, as indicated by the
detection of SiO(5-4).  In these cases, the Class I masers are
coincident with outflow lobes, and appear as clear cases of excitation
by active outflows.  In contrast, the third Class I source is
associated with an ultracompact HII region, and \emph{not} with Class
II masers.  The lack of SiO emission suggests the $^{13}$CO outflow is
a relic, consistent with its longer dynamical timescale.  Our data
show that massive young stellar objects associated only with Class I
masers are not necessarily young, and provide the first unambiguous
evidence that Class I masers may be excited by both young (hot core)
and older (UC HII) MYSOs within the same protocluster.

\end{abstract}


\section{Introduction}\label{intro}

The lack of a detailed, observationally based evolutionary sequence
for massive young stellar objects (MYSOs) limits our understanding of
the early stages of high mass (M$_{\rm ZAMS}>$8 M$_{\odot}$) star
formation.  Massive stars form in (proto)clusters, with younger
sources (such as hot cores) often found in close proximity to
ultracompact (UC) HII regions \citep[e.g.][and references
  therein]{Hunter06,Cyganowski07}.  Because massive protoclusters are
generally distant (D$>$1 kpc) and deeply embedded, most efforts to
develop evolutionary sequences have focused on cm-wavelength maser
transitions, which are amenable to high resolution study and
unaffected by extinction.  Four types of masers are ubiquitous: Class
I and II CH$_{3}$OH, H$_{2}$O, and OH.  Class I CH$_{3}$OH and
H$_{2}$O masers are collisionally pumped in shocked gas, while Class
II CH$_{3}$OH and OH masers are radiatively pumped by infrared
emission from warm dust \citep[e.g.][]{Elitzur89,Cragg02,Voronkov06}.
CH$_{3}$OH masers are key to proposed evolutionary sequences of masers
in MYSOs, which posit that Class I CH$_{3}$OH masers appear first
\citep[e.g.][]{Ellingsen07,Breen10evol}, based in part on their
observed association with outflows
\citep[e.g.][]{PlambeckMenten90,Kurtz04,maserpap}.  Class I-only
sources have thus generally been interpreted as tracing the earliest
stages of massive star formation \citep{Ellingsen06}.
Recent work, however, suggests that Class I CH$_{3}$OH masers may also 
be excited in shocks driven by expanding HII regions 
\citep{Voronkov10}, raising the possibility of multiple distinct
epochs of Class I maser activity during MYSO evolution
\citep{Chen11,Voronkov12}.  The prevalence of Class I-only sources is
currently unknown, primarily due to the lack of substantial untargeted
searches in Class I maser transitions \citep[see also][]{Voronkov10}.

\begin{figure*}
\plotone{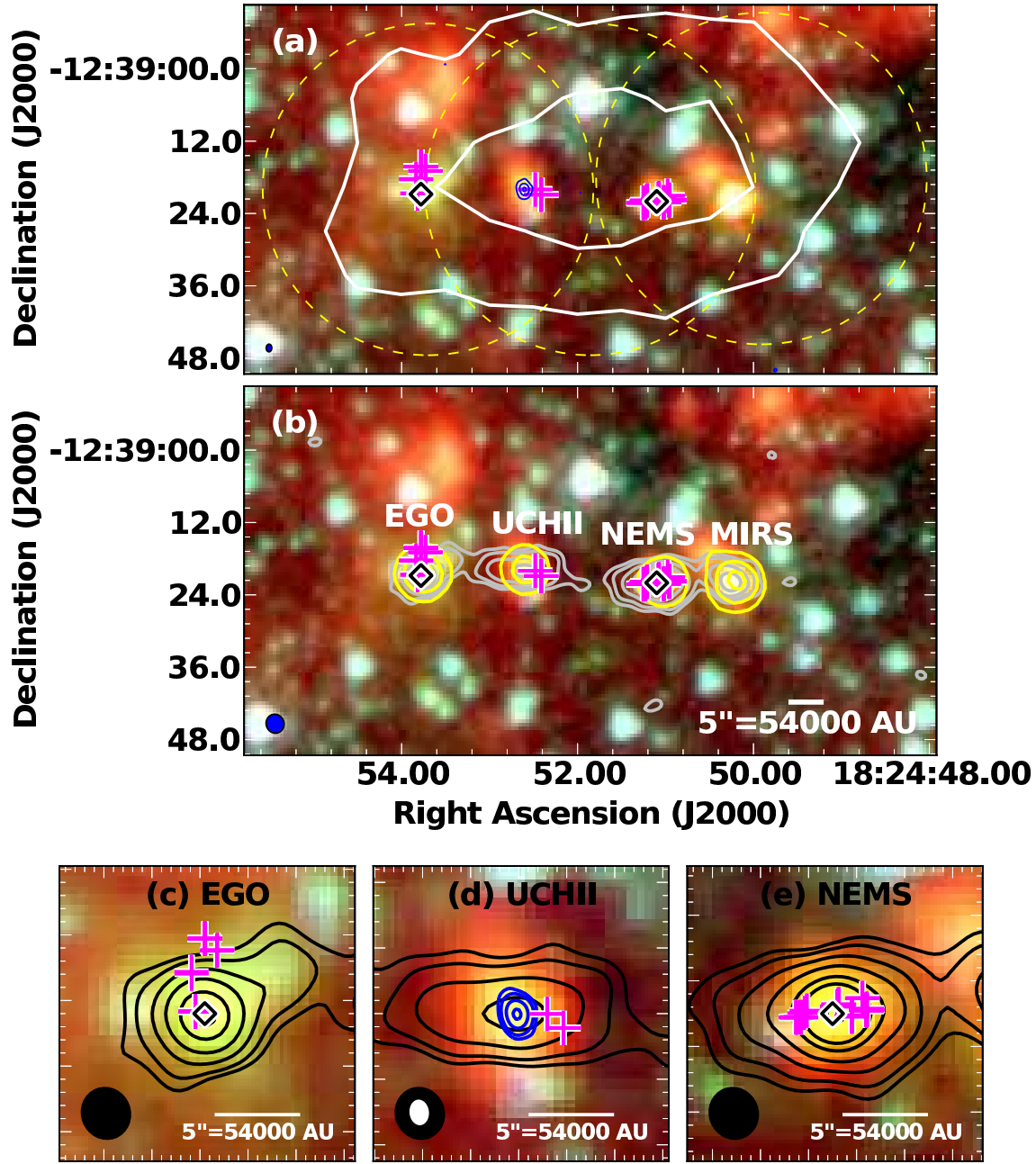}
\caption{\scriptsize Three color \emph{Spitzer} image (RGB: 8.0,4.5,3.6 $\mu$m)
of G18.67+0.03; positions of CH$_{3}$OH masers from \citet{maserpap}
are marked with $\Diamond$ (Class II) and $+$ (Class I).  In all
panels, the synthesized beam(s) are shown at lower left.  (a) SMA
pointings (dashed yellow circles) and contours of BGPS 1.1 mm emission
(white) and VLA 3.6 cm emission \citep[blue;][]{C11vla} are overlaid.
Contour levels: BGPS: (10,20)$\times\sigma, \sigma=$0.03 Jy beam$^{-1}$;
VLA: (4,40,120)$\times\sigma, \sigma=$3.06E-5 Jy beam$^{-1}$. (b) Contours
of SMA 1.3 mm continuum emission (gray) and MIPS 24 $\mu$m emission
(yellow). Levels: SMA:
(3,5,10,20,40,60,120)$\times\sigma, \sigma=$0.00145 Jy beam$^{-1}$.  MIPS:
600,1200,1800 MJy sr$^{-1}$.  (c-e): Zoom views of maser sources, with
SMA contours (black) as in (b).  In (d), VLA 1.2
cm continuum contours (blue) are also overlaid (levels (5,10,25,50)$\times\sigma, \sigma=$7.4E-5 Jy beam$^{-1}$).}
\label{intro_fig}
\end{figure*}

Our studies of a new sample of MYSO outflow candidates \citep[Extended
  Green Objects (EGOs), selected based on extended 4.5 $\mu$m emission
  in GLIMPSE images;][]{egocat} have identified a massive protocluster
\citep[M$_{\rm clump}\sim$3510 M$_{\odot}$;][]{atlasgal}
that provides a unique opportunity to understand the evolution
of Class I CH$_{3}$OH masers: G18.67+0.03.  \citet{maserpap} detected
44 GHz Class I CH$_{3}$OH masers towards three mid-infrared (MIR)
sources in this region: two (including the EGO) are also associated
with 6.7 GHz Class II CH$_{3}$OH masers (Fig.~\ref{intro_fig}).  The
third--the only source in the region with a cm-wavelength continuum
counterpart in deep Very Large Array (VLA) images \citep{C11vla}--has \emph{no} Class
II masers.  
\citet{Green11} assign both 6.7 GHz masers in G18.67+0.03 the far distance in their
HI self-absorption study of Class II masers.
Using the revised
kinematic distance prescription of \citet{Reid09} and thermal NH$_{3}$
data, \citet{nobeyama} find a similar far distance of 10.8 kpc,
which we adopt here.
 
In this Letter, we present a high-resolution Submillimeter Array
(SMA)\footnote{The Submillimeter Array is a joint project between the
Smithsonian Astrophysical Observatory and the Academia Sinica
Institute of Astronomy and Astrophysics and is funded by the
Smithsonian Institution and the Academia Sinica.} study of the
mm-wavelength dust continuum and molecular line emission toward
G18.67+0.03, to constrain the evolutionary states of the maser sources
and the presence/absence of active outflows.  
The SMA is well-suited for deciphering MYSO relative evolutionary states 
\citep{Beuther09,Zhang09,Wang12}.  
To provide kinematic
information for the ionized gas, we include complementary 1.3 cm Karl
G. Jansky Very Large Array (VLA) hydrogen recombination line and cm continuum
observations.

\section{Observations}\label{obs}

SMA observations at 1.3mm were obtained on 18 June 2010 with 7
antennas in compact configuration, in good weather
($\tau_{225GHz}\sim$0.07).  A three-pointing mosaic
was used to cover the extent of the (sub)mm clump
(Fig.~\ref{intro_fig}).  The primary beam (FWHP) at 1.3 mm is
$\sim$55\arcsec. The largest angular scale recoverable from these data
is $\sim$20\arcsec.  We observed $\sim$216.8-220.8 GHz in the lower
sideband (LSB) and $\sim$228.8-232.8 GHz in the upper sideband (USB),
each divided into 2$\times$2 GHz IFs, with a uniform channel width of
0.8125 MHz.  The calibrators were J1733-130 and J1911-201 (complex gain),
3C454.3 (bandpass), and Neptune (absolute flux density).  The LSB was flux-calibrated
using a model of Neptune's brightness distribution and the MIRIAD task
smaflux.  This direct approach cannot be used for the USB because of
strong $^{12}$CO in Neptune's atmosphere.  Instead, the USB flux density calibration was bootstrapped from the LSB assuming a spectral index
$\alpha=-$0.85 for 3C454.3 (based on SMA monitoring).  Comparison of
derived quasar flux densities
with SMA monitoring suggests the absolute flux density calibration is accurate
to better than $\sim$15\%.

\tabletypesize{\scriptsize}
\setlength{\tabcolsep}{1.5pt}
\begin{deluxetable*}{lcccccccccc}
\tablewidth{0pt}
\tablecaption{Properties of Millimeter Continuum Sources \label{cont_table}}
\tablehead{
\colhead{} & 
\multicolumn{4}{c}{Observed Properties} &
\multicolumn{5}{c}{Derived Properties}\\
\colhead{Source\tablenotemark{a}} &
\multicolumn{2}{c}{J2000 Coordinates\tablenotemark{b}} &
\colhead{Peak Intensity} & 
\colhead{Integ. Flux Density\tablenotemark{c}} &
\colhead{T$_{\rm dust}$} &
\colhead{$\tau_{\rm dust}$} & 
\colhead{M$_{\rm gas}$} & 
\colhead{N$_{H_{2}}$\tablenotemark{d}} & 
\colhead{n$_{H_{2}}$\tablenotemark{d}} &
\colhead{v$_{\rm LSR}$\tablenotemark{e}}\\
\colhead{Name} & 
\colhead{$\alpha$ ($^{\rm h}~~^{\rm m}~~^{\rm s}$)} &
\colhead{$\delta$ ($^{\circ}~~{\arcmin}~~{\arcsec}$)} &
\colhead{(mJy beam$^{-1}$)} & 
\colhead{(mJy)} & 
\colhead{(K)} &
\colhead{} &
\colhead{(M$_{\odot}$)} &
\colhead{(10$^{23}$ cm$^{-2}$)} &
\colhead{(10$^{5}$ cm$^{-3}$)} &
\colhead{km s$^{-1}$}
}
\tablecolumns{11}  
\startdata 
EGO & 18 24 53.77 & -12 39 20.8 & 96 & 153(36) & 175 & 0.003 & 30 & 0.7 & 2.8 & 79\\
CM1-UCHII\tablenotemark{f} & 18 24 52.59 & -12 39 20.0 & 36 & 126(41)& 50 & 0.008 & 100 & 2.1 & 8.6 & 80\\ 
NEMS & 18 24 51.09 & -12 39 22.0 & 187 & 347(44) & 185 & 0.006 & 70 & 1.5 & 6.1 & 82 \\ 
MIRS & 18 24 50.27 & -12 39 22.0 & 44 & 92(32) & & & & & &81\\ 
\\
Total & & & & 718\\ 
\enddata 
\tablenotetext{a}{As designated in \S\ref{results}.} 
\tablenotetext{b}{Coordinates of 1.3 mm continuum peak.  The number of significant figures reflects a one-pixel uncertainty.}
\tablenotetext{c}{Integrated flux density within 3$\sigma$ contour, measured using the CASAVIEWER program.  Following the approach developed for the CORNISH survey \citep{Purcell08}, we estimate uncertainties as  \begin{math}\sqrt{N_{\rm sr}\sigma_{\rm sky}^2(1+\frac{N_{\rm sr}}{N_{\rm sky}})}\end{math}, where N$_{\rm sr}$ is the number of pixels in the source aperture, N$_{\rm sky}$ is the number of pixels in an off-source ``sky'' aperture, and $\sigma_{\rm sky}$ is the standard deviation over the ``sky'' aperture.  The quoted uncertainties are averages for 3 choices of ``sky'' apertures.}  
\tablenotetext{d}{Beam-averaged} 
\tablenotetext{e}{Determined from compact molecular line emission, \S\ref{results}.}
\tablenotetext{f}{Physical properties calculated using the 1.3 mm flux density less the estimated free-free contribution of 5 mJy.}
 
\end{deluxetable*}

Initial calibration was performed in MIRIAD; the data were then
exported to CASA for further processing.  Each IF was processed
independently, with the continuum estimated using line-free channels
in the \emph{uv}-plane, and separated from the line emission.  The
continuum was self-calibrated and the resulting solutions applied
to the line data.  The data were imaged using Briggs
weighting and a robust parameter of 0.5.  The final combined continuum image
has a synthesized beamsize of 3\farcs16$\times$2\farcs95
(P.A.$=$21$^{\circ}$) and a 1$\sigma$ rms of 1.5 mJy beam$^{-1}$.
The line data were resampled to $\Delta$v=1.2 km s$^{-1}$, then
Hanning-smoothed; the typical rms per channel is $\sim$40 mJy beam$^{-1}$. 
All measurements were made from images corrected for the primary beam response.

The NRAO\footnote{The National Radio Astronomy Observatory is a
facility of the National Science Foundation operated under agreement
by the Associated Universities, Inc.} VLA observations of G18.67+0.03
were obtained in the C configuration on 7 January 2011 as part of a
1.3 cm line and continuum survey of massive protostellar objects
\citep[program AB1346;][]{rsro}.
The calibrators were J1832-1035 (complex gain),
J1924-2914 (bandpass), and 3C286 (absolute flux density).  A single pointing was
observed ($\alpha=$18$^{\rm h}$24$^{\rm m}$53$^{\rm s}$.700,
$\delta=-$12$^{\circ}$39\arcmin20\farcs0); the primary beam (FWHP) is
$\sim$2\arcmin.  The data were calibrated and imaged in CASA.  The
full VLA dataset will be described in a future
publication; here we consider only the 24.50878 GHz H64$\alpha$,
25.68788 GHz H63$\alpha$, and 24.64 GHz (1.2 cm) continuum data.  The self-calibrated continuum image
has a synthesized beamsize of 1\farcs46$\times$1\farcs08 (P.A.$=-$174$^{\circ}$) and a 1$\sigma$ rms of 74 $\mu$Jy beam$^{-1}$.
To improve S/N, we
smoothed the H63$\alpha$ and H64$\alpha$ data (to $\Delta$v$_{\rm
channel}=$4.8 km s$^{-1}$, $\theta_{beam}=$2\arcsec), then averaged
to obtain a final radio recombination line (RRL) image cube with $\sigma\sim$0.29 mJy beam$^{-1}$.  
All measurements were made
from images corrected for the primary beam response.

\section{Results}

\subsection{Millimeter Continuum and Line Emission}\label{results}

As shown in Figure~\ref{intro_fig}, we resolve four compact 1.3 mm
continuum sources, arranged in a line of $\sim$50\arcsec\/ (2.6 pc)
extent along the clump.  All have 24 $\mu$m counterparts.  For ease of
reference, we designate them (from E to W): EGO, CM1-UCHII, Non-EGO
maser source (NEMS), and mid-infrared source (MIRS).  CM1-UCHII refers
to the mm counterpart of the cm-wavelength continuum source designated
F G18.67+0.03-CM1 by \citet{C11vla}.  Observed millimeter continuum
properties are summarized in Table~\ref{cont_table}.  The SMA image
recovers 19$\pm$6\%
of the 1.1 mm BGPS flux density (3.8$^{+0.8}_{-0.7}$ Jy;
\citealt{Rosolowsky10}, corrected as per \citealt{Aguirre11}). 
The westernmost source (MIRS) falls outside the area searched for 44
GHz CH$_{3}$OH masers by \citet{maserpap}, so we do not discuss it
further in this Letter.

\begin{figure*}
\plotone{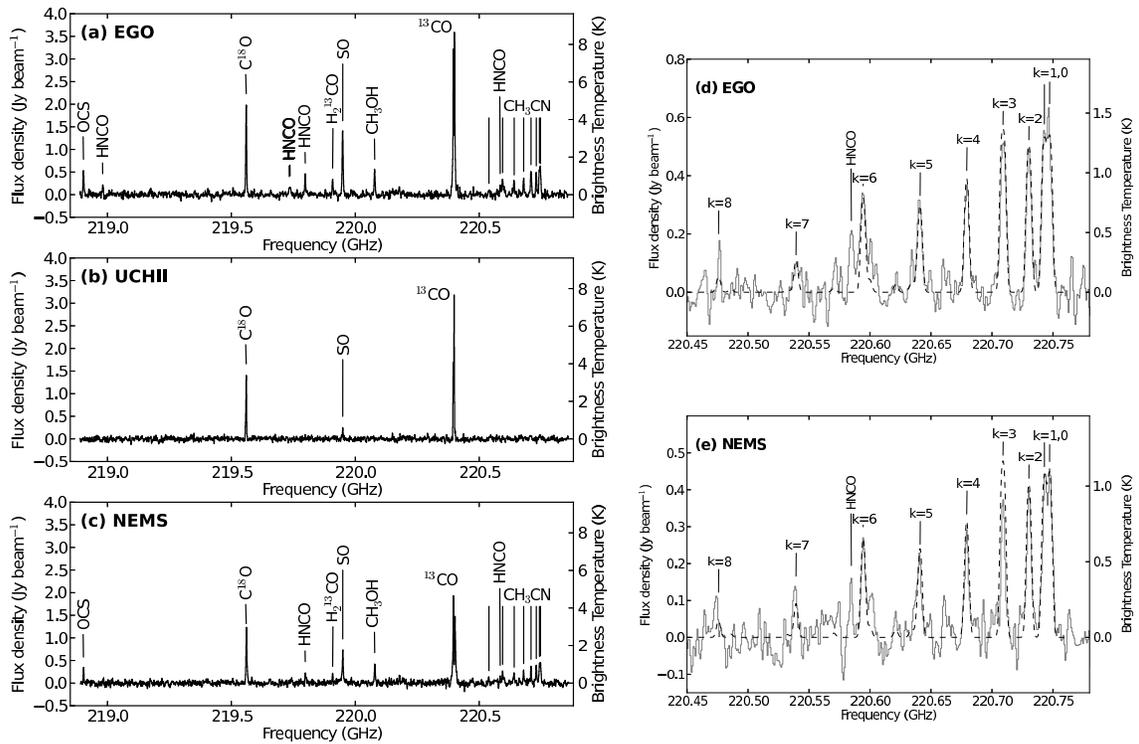}
\caption{SMA spectra towards the (a) EGO, (b) CM1-UCHII, and (c) NEMS 1.3 mm continuum peaks, showing 2 GHz of the LSB.  (d-e) SMA CH$_{3}$CN spectra (gray) overlaid with the best-fit model (dashed black line).}
\label{cont_peak_spectra_fig}
\end{figure*}

\begin{figure*}
\plotone{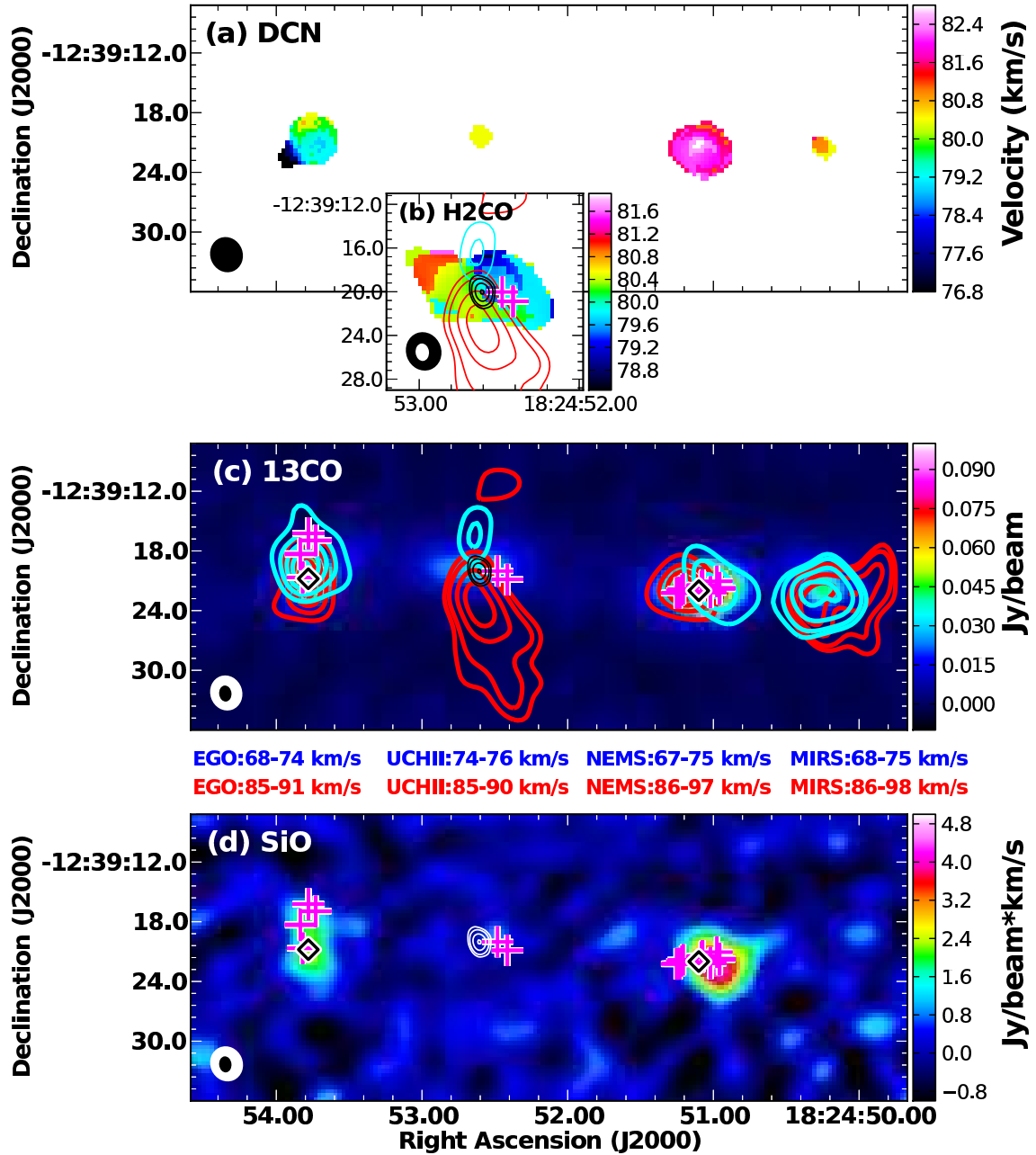}
\caption{\scriptsize SMA molecular gas kinematics.  
(a) DCN(3-2) first moment map.  (b) Zoom view of CM1-UCHII
  H$_{2}$CO(3$_{\rm0,3}$-2$_{\rm0,2}$) velocity field (moment 1,
  colorscale), overlaid with contours of VLA 1.2 cm continuum emission
  (black) and blue/red-shifted $^{13}$CO(2-1) emission.  (c) SMA 1.3
  mm continuum (colorscale) overlaid with contours of VLA 1.2 cm
  continuum (black) and composite maps of blue/red-shifted
  $^{13}$CO(2-1) emission.  $^{13}$CO is integrated over the indicated
  velocity ranges, optimized for each source.  Contour levels (Jy
  beam$^{-1}$ km s$^{-1}$): EGO: blue: 1,2,3,4,5; red:
  1,2,3,4. CM1-UCHII: blue: 2,3; red: 2,3,5,7. NEMS: blue: 2,4,6; red:
  2,4,6,8. MIRS: blue: 1.5,2,3,4; red: 1.5,2,3,4.  (d) SiO(5-4) integrated
  intensity map (moment 0, colorscale), overlaid with VLA 1.2 cm
  continuum contours (white).  VLA contour levels
  as in Figure~\ref{intro_fig}.  Positions of CH$_{3}$OH masers from
  \citet{maserpap} are marked with $\Diamond$ (Class II) and $+$
  (Class I).  Synthesized beam(s) are shown at lower left in each
  panel.}
\label{kinematics_fig}
\end{figure*}

Only a few species (including CO isotopes, SO, DCN, and low-excitation
lines of H$_{2}$CO and CH$_{3}$OH) are detected towards all three
maser sources.  The LSR velocities of the sources, determined from
compact molecular line emission, are within 3 km s$^{-1}$ of each other
(Fig.~\ref{kinematics_fig}a, Table~\ref{cont_table}).  
Figure~\ref{cont_peak_spectra_fig}a-c presents a comparison of the
spectra at the EGO, CM1-UCHII, and NEMS 1.3 mm continuum peaks across
2 GHz of the SMA band.  The EGO and NEMS both exhibit copious
molecular line emission from classic hot core tracers, such as
CH$_{3}$CN, OCS, and high-excitation CH$_{3}$OH.  In contrast,
CM1-UCHII is line-poor: none of these tracers are detected.

All three Class I CH$_{3}$OH maser sources are associated with both
redshifted and blueshifted $^{13}$CO(2-1) emission, with velocity
extents of 23, 16, and 30 km s$^{-1}$ for the EGO, CM1-UCHII, and
NEMS, respectively (Fig.~\ref{kinematics_fig}c).  In all cases, there
is a clear spatial offset between the red and blue lobes, which are
centered on a compact source.
These characteristics indicate that the high-velocity $^{13}$CO(2-1)
emission traces bipolar molecular outflows.  The 44 GHz masers
associated with the EGO and NEMS are coincident with outflow lobes.
The CM1-UCHII 44 GHz masers are located, in projection, near the edge
of its redshifted lobe.  In addition to the N-S outflow, there is
evidence for an E-W gradient in low-velocity molecular gas near the
UCHII region (v$-$v$_{\rm systemic}\lesssim$1.5 km s$^{-1}$;
Fig.~\ref{kinematics_fig}b).  This E-W velocity gradient could
indicate a second, slower outflow, and/or large-scale rotation or
infall.  Importantly, the outflows associated with the EGO and NEMS
are detected in SiO(5-4) emission, while the outflow(s) associated
with CM1-UCHII is not (Fig.~\ref{kinematics_fig}d).
  
\subsection{Temperatures and Dust Masses}\label{results_temps}

We estimate gas temperatures for the EGO and NEMS by fitting the
J=12-11 CH$_{3}$CN ladder using the method described in \citet{C11}.
In brief, the model accounts for optical depth effects and emission
from the CH$_{3}^{13}$CN isotope: the temperature, source
size(diameter), and CH$_{3}$CN column density and linewidth are free
parameters.  Figure~\ref{cont_peak_spectra_fig}d-e show the best-fit models, overlaid on the
observed spectra.  The best-fit temperatures, source sizes, and
CH$_{3}$CN column densities are quite similar for the two sources:
T=175K and 185K, log[N$_{\rm CH_{3}CN}$(cm$^{-2}$)]=16.66 and 16.58, and
d$_{\rm source}$=0\farcs45 (4900 AU) and 0\farcs39 (4200 AU) for the
EGO and NEMS, respectively. 

Gas masses and (beam-averaged) column and number densities estimated
from the 1.3 mm thermal dust emission, using the methodology of
\citet{C11}, are presented in Table~\ref{cont_table}.  The mass
estimates for the EGO and NEMS assuming T$_{\rm dust}$=T$_{\rm
CH_{3}CN}$ ($\sim$30 and 70 M$_{\odot}$, respectively) are likely lower limits; the CH$_{3}$CN-emitting regions
are unresolved at the scale of our SMA observations ($\theta_{\rm
syn.}\sim$33,000 AU), while the 1.3 mm continuum emission appears
somewhat more extended.  
For comparison, the gas masses of the CH$_{3}$CN-emitting
regions, estimated from the best-fit source sizes and column
densities, are $\sim$40 and 30 M$_{\odot}$ for the EGO and NEMS, for
CH$_{3}$CN/H$_{2}=$10$^{-8}$.  Since the CH$_{3}$CN/H$_{2}$
abundance in hot cores is uncertain, and values an order of magnitude
higher/lower are plausible \citep[see][]{C11}, this is reasonable agreement.

Constraining the gas temperature of CM1-UCHII is more difficult
because of the paucity of associated line emission: no CH$_{3}$CN is
detected, and only one CH$_{3}$OH line.  Most lines detected have
E$_{\rm upper}<$50 K; the highest-excitation transitions have E$_{\rm
  upper}\sim$68 K (H$_{2}$CO).  In addition, while the ionized gas is
unresolved (\S\ref{ion_results}), the mm-wavelength continuum emission
is extended E-W, with scale $>$0.5 pc.  For the estimates in
Table~\ref{cont_table}, we adopt a temperature of 50 K, and subtract
the free-free contribution to the 1.3 mm flux density (extrapolated
from the 1.2 cm integrated flux density assuming optically thin
emission, S$_{\nu}\propto\nu^{-0.1}$).

\newpage

\subsection{Ionized Gas Properties}\label{ion_results}

We detect the H63$\alpha$ and H64$\alpha$ recombination lines towards
G18.67+0.03-CM1 \citep{C11vla}.  The fitted RRL velocity (77.3$\pm$0.9
km s$^{-1}$) is consistent with the molecular-gas velocity.  From the
line-to-continuum ratio of 0.48, the fitted FWHM linewidth of 27.0 km
s$^{-1}$, and the fitted source size of 0\farcs63 (from the continuum
image), we derive an electron temperature of $\sim$5100 K and density
of $\sim$1.9$\times$10$^{4}$ cm$^{-3}$ \citep[following the method of
  e.g.][]{Garay86,Sewilo11}.  If bulk motions (e.g.\ expansion)
dominate the linewidth, this could explain the low derived T$_{e}$;
however, the RRL emission is too weak
to investigate this possibility.  The cm-wavelength spectral index is
consistent with optically thin free-free emission, and the ionizing
photon flux \citep[$\sim$1$\times$10$^{47}$ s$^{-1}$, calculated as
  in][]{C11vla} corresponds to a single ionizing star of spectral type
B0.5V \citep{Smith02}.

\newpage
\section{Discussion}\label{discussion}

\subsection{Evolutionary States of Class I Maser Driving Sources}

All three maser sources in G18.67+0.03 are associated with massive
($\gtrsim$30 M$_{\odot}$), dense (n$_{\rm H_{2}}>$10$^{5}$ cm$^{-3}$)
millimeter continuum cores, and with $^{13}$CO outflows.  CO (and
HCO$^{+}$) may trace relic outflows from MYSOs
\citep[e.g.][]{Klaassen06,Klaassen07,Hunter08}.  SiO emission,
however, specifically indicates recent shocks and \emph{active}
outflows, since the gas-phase SiO abundance is enhanced by shocks and
remains so for only $\sim$10$^{4}$ years \citep[e.g.][]{PdF97}.  The
EGO and NEMS, which have \emph{both} Class I and II CH$_{3}$OH masers,
thus share two other key characteristics: (1) their outflows are
\emph{active}, as evidenced by SiO(5-4) emission; and (2) the driving
sources exhibit hot core line emission (e.g.\ CH$_{3}$CN, CH$_{3}$OH,
OCS).  In summary, the Class I masers in these sources appear as clear
cases of excitation by outflows.

In contrast, the Class I-only maser source lacks indicators of youth:
neither SiO nor hot core line emission is detected in our
observations.  The 44 GHz masers lie near a cm continuum source,
detected in RRLs; its derived ionized gas properties are consistent
with a UCHII region.  Two mechanisms have been posited that could
explain Class I maser emission associated with such a comparatively
evolved source.  \citet{Voronkov06} suggested that Class I masers
might be long-lived, persisting after the exciting shock had
dissipated.
More recently, \citet{Voronkov10} noted the
proximity of cm continuum emission to three of the four known examples
of 9.9 GHz Class I CH$_{3}$OH masers.  They suggested that shocks
driven into the surrounding cloud by expanding HII regions were
responsible for these masers, and that this mechanism should
also apply to other Class I transitions.  These two scenarios
have somewhat different implications for maser evolutionary sequences:
the first implies that Class I masers \emph{outlast} the Class II
phase; the second, that Class I masers may appear \emph{more than
once}, excited by young \emph{or} evolved MYSOs \citep[see also][]{Voronkov12}.

In G18.67+0.03, the lack of associated SiO emission suggests that the
CM1-UCHII $^{13}$CO outflow is a relic.  It also has a smaller
velocity extent than the EGO and NEMS $^{13}$CO outflows, despite
being more spatially extended: additional evidence that the CM1-UCHII
outflow is older (t$_{\rm dyn}\sim$ 19000, 26000, and 66000 for the
NEMS, EGO, and CM1-UCHII respectively).
In this picture, the CM1-UCHII $^{13}$CO outflow would have been
driven by the now-ionizing source, prior to the creation of the UCHII
region \citep[analogous to the E-W outflow in
  G5.89-0.39;][]{Hunter08}.  The Class I masers are near the edge of
the redshifted lobe, consistent with an association with the relic(?)
flow, and so supporting the scenario in which long-lasting Class I
maser activity extends beyond the Class II maser lifetime.  This
interpretation, however, is not conclusive; the Class I masers are
also near the UCHII region, and we cannot rule out their being excited
by HII region-driven shocks.
In either case, these Class I masers are clearly associated with a more evolved MYSO.

\subsection{Class I Maser Excitation as a Key?}

\begin{figure}
\begin{center}
\plotone{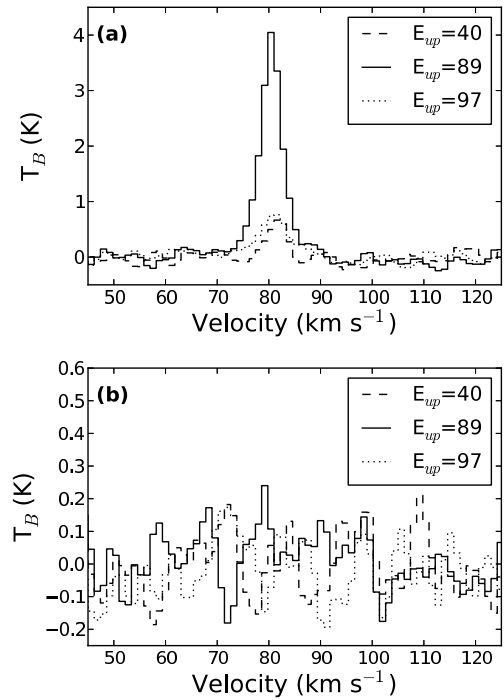}
\end{center}
\caption{SMA CH$_{3}$OH spectra at the position of the strongest 44 GHz maser
  associated with (a) NEMS and (b) CM1-UCHII.  Solid line: 8$_{-1,8}$-7$_{0,7}$ (229.759 GHz); dashed
  line: 3$_{-2,2}$-4$_{-1,4}$ (230.027 GHz); dotted line:
  8$_{0,8}$-7$_{1,6}$ (220.078 GHz).}
\label{mm_maser_spec_fig}
\end{figure}

Probable maser emission in the Class I 229.759 GHz CH$_{3}$OH(8$_{\rm
  -1,8}$-7$_{\rm 0,7}$)E transition is a conspicuous feature of recent
SMA observations of MYSO outflows \citep[e.g.][and references
  therein]{C11}.  In these outflows, 229 GHz emission is generally
colocated (spatially and spectrally) with lower frequency Class I
masers, and the brightest features in all transitions coincide
\citep[see also][]{Fish11}.  In G18.67+0.03, we detect strong 229.759
GHz emission coincident with the brightest NEMS 44 GHz masers
(Fig.~\ref{mm_maser_spec_fig})--those associated with the blueshifted,
SiO-rich western outflow lobe.  Like many previous 229 GHz studies,
our beam is too large to establish masing based on brightness
temperature.  \citet{Slysh02} proposed the 229.759/230.027 ratio as a
diagnostic, with values $>$3 indicating nonthermal 229 GHz emission:
the ratio for the NEMS position shown in
Figure~\ref{mm_maser_spec_fig} is 5.5.  This NEMS 44 GHz maser is the
second-brightest detected by \citet{maserpap} in G18.67+0.03; the
brightest is the western of the two associated with CM1-UCHII
(Fig.~\ref{intro_fig}d; I$_{\rm peak,44GHz}\sim$2 Jy beam$^{-1}$).  No
229.759 GHz emission is detected at this position
(Fig.~\ref{mm_maser_spec_fig}).  The 44 GHz masers associated with the
EGO are weak ($<$ 0.5 Jy beam$^{-1}$), and the mm line ratios are
ambiguous.

These results suggest that excitation conditions may differ
in older and younger Class I maser sources in a manner that produces
observable differences in the relative strengths of various Class
I maser transitions.  Detailed comparisons of the 44 and 229 GHz maser
spots are, however, limited by the relatively coarse spatial and
spectral resolution of the current SMA data.  In general, additional
modeling and observations are needed to understand what information
about physical conditions and evolutionary state can be extracted from
the presence/absence of different Class I CH$_{3}$OH masers,
particularly for mm-wavelength transitions \citep[see also discussion
in][]{Fish11}.  For G18.67+0.03, higher angular resolution
observations are needed to confirm the maser nature of the 229 GHz
emission, and to spatially resolve the potential mm masers from the
hot cores in the EGO and NEMS; high resolution 36 GHz observations may
also be helpful \citep[e.g.][]{Voronkov12}.

\section{Summary}\label{conclusions}

Our high resolution SMA and VLA observations provide the first
unambiguous evidence of Class I CH$_{3}$OH masers being excited by
both young and more evolved MYSOs \emph{within the same protocluster}.
The two hot cores, also associated with Class II CH$_{3}$OH masers,
drive active outflows traced by SiO(5-4) emission and Class I masers.
In contrast, the UCHII region is associated only with Class I
CH$_{3}$OH masers and with an older (possibly relic) $^{13}$CO
outflow; the UCHII shows neither SiO nor hot core line emission.
These results further demonstrate the limitations of current
evolutionary sequences for maser emission \citep[see
also][]{Voronkov10,Chen11}.  In particular, our data show that MYSOs
associated \emph{only with Class I CH$_{3}$OH masers (no Class II)}
are not necessarily young, in contrast to published sequences.  This
work highlights the importance of high angular resolution
multiwavelength observations for constraining MYSO evolutionary
states, and disentangling an observation-based MYSO evolutionary
sequence.

\acknowledgments

This research made use of NASA's Astrophysics Data System
Bibliographic Services,
APLpy, an open-source plotting package for Python hosted at
http://aplpy.github.com, and the myXCLASS program
(http://www.astro.uni-koeln.de/projects/schilke/XCLASS), which
accesses the CDMS (http://www.cdms.de) and JPL
(http://spec.jpl.nasa.gov) molecular databases.  C.J. Cyganowski is supported
by an NSF AAPF under award AST-1003134.


\begin{thebibliography}

\bibitem[Aguirre et al.(2011)]{Aguirre11} Aguirre, J.~E., 
Ginsburg, A.~G., Dunham, M.~K., et al.\ 2011, \apjs, 192, 4 

\bibitem[Beuther et al.(2009)]{Beuther09} Beuther, H., Zhang, Q., 
Bergin, E.~A., \& Sridharan, T.~K.\ 2009, \aj, 137, 406 


\bibitem[Breen et al.(2010a)]{Breen10evol} Breen, S.~L., Ellingsen, 
S.~P., Caswell, J.~L., \& Lewis, B.~E.\ 2010, \mnras, 401, 2219 

\bibitem[Brogan et al.(2011)]{rsro} Brogan, C.~L., Hunter, 
T.~R., Cyganowski, C.~J., et al.\ 2011, \apjl, 739, L16 




\bibitem[Chen et al.(2011)]{Chen11} Chen, X., Ellingsen, 
S.~P., Shen, Z.-Q., Titmarsh, A., \& Gan, C.-G.\ 2011, \apjs, 196, 9 



\bibitem[Cragg et al.(2002)]{Cragg02} Cragg, D.~M., Sobolev, 
A.~M., \& Godfrey, P.~D.\ 2002, \mnras, 331, 521 

\bibitem[Cyganowski et al.(2007)]{Cyganowski07} Cyganowski, C.~J., 
Brogan, C.~L., \& Hunter, T.~R.\ 2007, \aj, 134, 346 


\bibitem[Cyganowski et al.(2008)]{egocat}
Cyganowski, C. J., Whitney, B.A., Holden, E., Braden, E., Brogan, C.L.,
Churchwell, E., Indebetouw, R., Watson, D.F., Babler, B.L., Benjamin, R.,
Gomez, M., Meade, M.R., Povich, M.S. Robitaille, T.P., \& Watson, C.\ 2008,
\aj, 136, 2391

\bibitem[Cyganowski et al.(2009)]{maserpap} Cyganowski, C.~J., 
Brogan, C.~L., Hunter, T.~R., \& Churchwell, E.\ 2009, \apj, 702, 1615 

\bibitem[Cyganowski et al.(2011a)]{C11} Cyganowski, C.~J., 
Brogan, C.~L., Hunter, T.~R., Churchwell, E., 
\& Zhang, Q.\ 2011, \apj, 729, 124 

\bibitem[Cyganowski et al.(2011b)]{C11vla} Cyganowski, C.~J., 
Brogan, C.~L., Hunter, T.~R., \& Churchwell, E.\ 2011, \apj, 743, 56 

\bibitem[Cyganowski et al.\ (2012)]{nobeyama} Cyganowski, C.~J., et al., ApJ, submitted.

\bibitem[Elitzur et al.(1989)]{Elitzur89} Elitzur, M., 
Hollenbach, D.~J., \& McKee, C.~F.\ 1989, \apj, 346, 983 

\bibitem[Ellingsen et al.(2007)]{Ellingsen07} Ellingsen, S.~P., 
Voronkov, M.~A., Cragg, D.~M., Sobolev, A.~M., Breen, S.~L., 
\& Godfrey, P.~D.\ 2007, IAU Symposium, 242, 213 

\bibitem[Ellingsen(2006)]{Ellingsen06} Ellingsen, S.~P.\ 2006, 
\apj, 638, 241 

\bibitem[Fish et al.(2011)]{Fish11} Fish, V.~L., Muehlbrad, 
T.~C., Pratap, P., et al.\ 2011, \apj, 729, 14 


\bibitem[Garay et al.(1986)]{Garay86} Garay, G., Rodriguez, 
L.~F., \& van Gorkom, J.~H.\ 1986, \apj, 309, 553 


\bibitem[Green 
\& McClure-Griffiths(2011)]{Green11} Green, J.~A., \& McClure-Griffiths, N.~M.\ 2011, \mnras, 417, 2500 

\bibitem[Hunter et al.(2008)]{Hunter08} Hunter, T.~R., Brogan, 
C.~L., Indebetouw, R., \& Cyganowski, C.~J.\ 2008, \apj, 680, 1271 




\bibitem[Hunter et al.(2006)]{Hunter06} Hunter, T.~R., Brogan, C.~L., Megeath,
S.~T., Menten, K.~M., Beuther, H., \& Thorwirth, S.\ 2006, \apj, 649, 888

\bibitem[Klaassen 
\& Wilson(2007)]{Klaassen07} Klaassen, P.~D., \& Wilson, C.~D.\ 2007, \apj, 663, 1092 




\bibitem[Klaassen et al.(2006)]{Klaassen06} Klaassen, P.~D., 
Plume, R., Ouyed, R., von Benda-Beckmann, A.~M., 
\& Di Francesco, J.\ 2006, \apj, 648, 1079 




\bibitem[Kurtz et al.(2004)]{Kurtz04} Kurtz, S., Hofner, P., \& {\'A}lvarez,
C.~V.\ 2004, \apjs, 155, 149

\bibitem[Pineau des Forets et al.(1997)]{PdF97} Pineau des 
Forets, G., Flower, D.~R., 
\& Chieze, J.-P.\ 1997, Herbig-Haro Flows and the Birth of Stars, 182, 199 




\bibitem[Plambeck \& Menten(1990)]{PlambeckMenten90} Plambeck, R.~L., \&
Menten, K.~M.\ 1990, \apj, 364, 555

\bibitem[Purcell et al.(2008)]{Purcell08} Purcell, C.~R., Hoare, M.~G., 
\& Diamond, P.\ 2008, Massive Star Formation: Observations Confront Theory, 387, 389 

\bibitem[Reid et al.(2009)]{Reid09} Reid, M.~J., et al.\ 2009, 
\apj, 700, 137 

\bibitem[Rosolowsky et al.(2010)]{Rosolowsky10} Rosolowsky, E., et al.\ 2010, \apjs, 188, 123 

\bibitem[Schuller et al.(2009)]{atlasgal} Schuller, F., et al.\ 2009, \aap, 504, 415 

\bibitem[Sewi{\l}o et al.(2011)]{Sewilo11} Sewi{\l}o, M., 
Churchwell, E., Kurtz, S., Goss, W.~M., \& Hofner, P.\ 2011, \apjs, 194, 44 

\bibitem[Slysh et al.(2002)]{Slysh02} Slysh, V.~I., Kalenski{\u 
i}, S.~V., \& Val'tts, I.~E.\ 2002, Astronomy Reports, 46, 49 



\bibitem[Smith et al.(2002)]{Smith02} Smith, L.~J., Norris, 
R.~P.~F., \& Crowther, P.~A.\ 2002, \mnras, 337, 1309 


\bibitem[Voronkov et al.(2006)]{Voronkov06} Voronkov, M.~A., 
Brooks, K.~J., Sobolev, A.~M., et al.\ 2006, \mnras, 373, 411 

\bibitem[Voronkov et al.(2010)]{Voronkov10} Voronkov, M.~A., 
Caswell, J.~L., Ellingsen, S.~P., 
\& Sobolev, A.~M.\ 2010, \mnras, 405, 2471 

\bibitem[Voronkov et al.(2012)]{Voronkov12} Voronkov, M.~A., 
Caswell, J.~L., Ellingsen, S.~P., et al.\ 2012, arXiv:1203.5492 

\bibitem[Wang et al.(2012)]{Wang12} Wang, Y., Beuther, H., 
Zhang, Q., et al.\ 2012, \apj, 754, 87 


\bibitem[Zhang et al.(2009)]{Zhang09} Zhang, Q., Wang, Y., 
Pillai, T., \& Rathborne, J.\ 2009, \apj, 696, 268 



\end{thebibliography}
\end{document}